\documentclass[aps,bibnotes,superscriptaddress,twocolumn]{revtex4}

\usepackage{graphicx} 
\usepackage[format=plain,justification=raggedright,singlelinecheck=false,font=small,labelfont=bf,labelsep=space]{caption} 
\usepackage{amssymb}
\usepackage{amsmath}
\usepackage{amsfonts}
\usepackage{hyperref}

\usepackage{subfigure}
\usepackage{upgreek}

\newcommand{\gammadot}{\ensuremath{\dot{\gamma}}}
\newcommand{\TBTT}{TB-TT transition}
\newcommand{\Ca}{\ensuremath{\text{Ca}}}
\newcommand{\Ht}{\ensuremath{\text{Ht}}}
\newcommand{\Cacrit}{\ensuremath{\text{Ca}^*_{\text{cr}}}}
\newcommand{\mods}{\ensuremath{\kappa_{\text{S}}}}
\newcommand{\modb}{\ensuremath{\kappa_{\text{B}}}}
\newcommand{\moda}{\ensuremath{\kappa_\alpha}}
\newcommand{\modA}{\ensuremath{\kappa_{\text{A}}}}
\newcommand{\modV}{\ensuremath{\kappa_{\text{V}}}}
\newcommand{\Uwall}{\ensuremath{u_{\mathrm{w}}}}
\providecommand{\st}[1]{_{\text{#1}}}

\usepackage{color}
\usepackage[normalem]{ulem}

\begin{document}
 \title{Crossover from tumbling to tank-treading-like motion in dense simulated suspensions of red blood cells}
 \author{Timm Kr\"uger}
\affiliation{Centre for Computational Science, University College London, 20 Gordon Street, London WC1H 0AJ, United Kingdom}
 \author{Markus Gross}
 \affiliation{Interdisciplinary Centre for Advanced Materials Simulation (ICAMS), Ruhr-Universit\"at Bochum, Universit\"atsstr. 150, 44780 Bochum, Germany}
  \author{Dierk Raabe}
 \affiliation{Max-Planck Institut f\"ur Eisenforschung, Max-Planck Str.~1, 40237 D\"usseldorf, Germany}
  \author{Fathollah Varnik}
 \affiliation{Interdisciplinary Centre for Advanced Materials Simulation (ICAMS), Ruhr-Universit\"at Bochum, Universit\"atsstr. 150, 44780 Bochum, Germany}
 \affiliation{Max-Planck Institut f\"ur Eisenforschung, Max-Planck Str.~1, 40237 D\"usseldorf, Germany}

\begin{abstract}
Via computer simulations, we provide evidence that the shear rate induced red blood cell tumbling-to-tank-treading transition also occurs at quite high volume fractions, where collective effects are important. The transition takes place as the ratio of effective suspension stress to the characteristic cell membrane stress exceeds a certain value and does not explicitly depend on volume fraction or cell deformability. This value coincides with that for a transition from an orientationally less ordered to a highly ordered phase. The average cell deformation does not show any signature of the transition, but rather follows a simple scaling law independent of volume fraction.\end{abstract}

\maketitle

\section{Introduction}

Despite the large interest in a better understanding of the circulatory system and related diseases, there are still many open issues regarding the microscopic mechanisms which ultimately determine the rheology of suspensions of red blood cells (RBCs), the major particulate constituent of blood. Focusing on purely mechanical aspects, a RBC can be considered as a thin and flexible incompressible biconcave membrane which encloses an internal fluid (haemoglobin solution) and resists shearing and bending.

Depending on the ambient shear rate $\gammadot$, viscosity contrast (ratio between internal and external fluid viscosities), membrane viscoelasticity and other parameters, one observes tumbling, swinging, or tank-treading motion of isolated RBCs \cite{schmid-schonbein_fluid_1969, skotheim2007red, abkarian2007swinging, sui_dynamic_2008}. While in the case of single vesicles the dynamical phase space has been investigated  thoroughly \cite{keller_motion_1982, beaucourt2004steady, noguchi_fluid_2004, kantsler2006transition, misbah2006vacillating, forsyth2011multiscale, yazdani2011tank}, there are only few studies of dense suspensions of deformable particles or RBCs \cite{doddi_three-dimensional_2009, clausen_aidun_jfm2011, fedosov_predicting_2011, reasor_aidun_jfm2013}. These studies investigate the influence of concentration, deformabilty and aggregation tendency on suspension rheology.

In the present work, we focus on a dynamic phenomenon in dense RBC suspensions. Via computer simulations, we provide evidence for the transition from a tumbling to a tank-treading-like state (henceforth called \TBTT) upon increasing the capillary number $\Ca$. The capillary number is defined as the ratio of fluid stress to the characteristic membrane stress, where the latter is related to the shear elasticity of the cell.
Transcending previous studies, we find that the \TBTT\ occurs not only in the well-known case of a dilute suspension, but up to haematocrit values as high as $\Ht=65\%$ (haematocrit, or volume fraction, is the ratio of the volume occupied by RBCs to the total suspension volume). It is shown that---for all studied shear rates, cell deformabilities, and volume fractions---this transition is characterised by an effective capillary number $\Ca^*$ (ratio between \emph{effective} suspension stress and the characteristic membrane stress) rather than by the bare capillary number $\text{Ca}$. A detailed analysis of the average RBC inclination angle $\bar \theta$ (a measure of average cell orientation) and the corresponding order parameter $Q_>$ is also provided. When plotted versus $\Ca^*$, all values for $Q_>$ and $\bar \theta$ tend towards a master curve as $\Ca^*$ exceeds a certain threshold value $\Cacrit$ which, remarkably, coincides with that for the onset of the \TBTT\ for a single cell.

Our results provide evidence that the change from tumbling to tank-treading-like dynamics is accompanied by a phase transition from weak to strong orientational ordering of the cells. Interestingly, the average deformation of a cell, quantified by the Taylor deformation parameter $D_{\text{a}}$ changes continuously in the investigated parameter range, without any signature of the observed transition.

The article is organised as follows. The setup of the simulations is discussed in section \ref{sec_sim}. Section \ref{sec_results} contains the results and discussions of the \TBTT\ (section \ref{subsec_tbtt}), the orientational order (section \ref{subsec_order}), and the RBC deformation (section \ref{subsec_deformation}). We conclude our work and provide an outlook in section \ref{sec_conclusions}.

\section{Simulation setup}
\label{sec_sim}

\begin{figure}
 \centering
 \subfigure[\label{subfig_snapshot} simulation snapshot]{\includegraphics[width=0.35\linewidth]{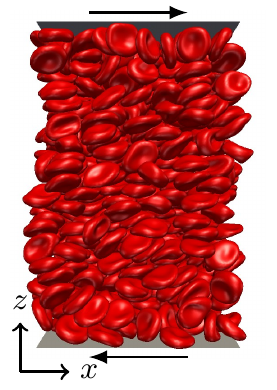}}
 \subfigure[\label{subfig_profiles} volume fraction, stress, and velocity profiles]{\includegraphics[width=0.625\linewidth]{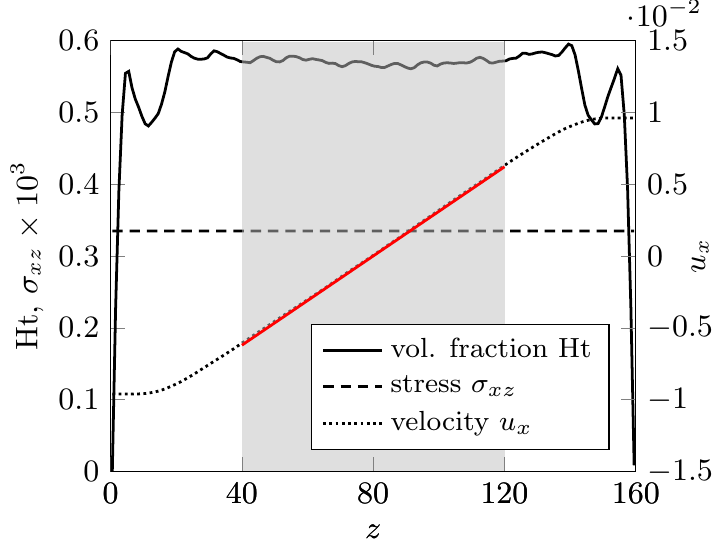}}
 \caption{\label{fig_density_and_stress}
 (a) Snapshot of a simulation with average haematocrit (volume fraction) $\text{Ht} = 55\%$ and capillary number $\Ca = 0.011$. The bottom (top) wall is located at $z=0$ ($z=L_z=160$) and moves to the left (right) with velocity $\pm \Uwall$ (black arrows). (b) Layer-resolved haematocrit, $\Ht$, shear stress, $\sigma_{xz}$, and flow velocity, $u_x$, for the same control parameters as in panel (a). Each curve is obtained as average over steady state and independent runs. In the central region (grey, from $L_z/4$ to $3L_z/4$), the volume fraction is nearly constant, the velocity profile is linear (red solid line), and the stress is constant.}
\end{figure}

\begin{figure}
 \centering
 \includegraphics[width=\linewidth]{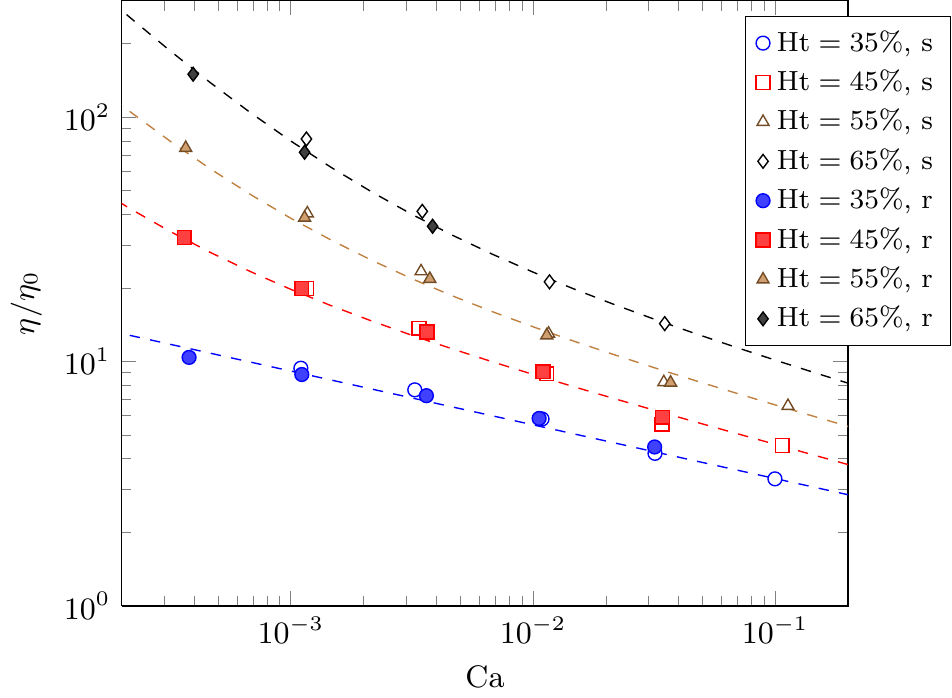}
 \caption{\label{fig_supp_viscosities} Relative suspension viscosity $\eta / \eta_0$ as function of bare capillary number $\text{Ca} = \eta_0 \dot{\gamma}r / \mods$. $\eta_0$ is the reference viscosity of the suspending fluid without red blood cells. For all volume fractions, the suspension is  shear thinning. Additionally, the viscosity increases with increasing volume fraction. One observes that the data for soft and hard RBCs overlap, when plotted versus Ca. Error bars are of the order of the symbol size. 
It is also noteworthy that the viscosities can be approximated by a Herschel-Bulkley law (dashed lines) of the form $\eta / \eta_0 = a + b \times \text{Ca}^c$ with the parameters $(a, b, c) = (0, 2.0, 0.78)$ for $35\%$, $(0.004, 2.4, 0.73)$ for $45\%$, $(0.015, 3.4, 0.72)$ for $55\%$, and $(0.043, 5.0, 0.71)$ for $65\%$.}
\end{figure}

\begin{figure}
 \centering
\includegraphics[width=0.6\linewidth]{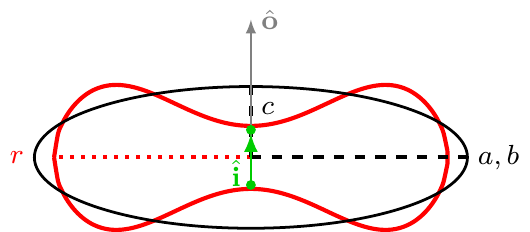}
 \caption{\label{fig_rbc_ellips}
 Cross-section through an undeformed red blood cell (red) together with its equivalent inertia ellipsoid (black). $r$ is the large RBC radius (dotted), $a,b$ are the two large semiaxes ($a=b=1.1r$) and $c$ is the small semiaxis ($c=0.36r$) of the inertia ellipsoid. The vector $\hat{\mathbf{o}}$ is perpendicular to the $a$-$b$-plane and is given by the normalized eigenvector of the inertia ellipsoid corresponding to the smallest eigenvector (i.e., the shortest semiaxis, $c$). The intrinsic membrane orientation vector $\hat{\mathbf{i}}$, on the other hand, is computed from the positions of the vertices of the RBC mesh and connects the centre of masses (green dots) of the top and bottom halfs of the undeformed membrane. The identity of a mesh vertex of being in the ``top'' or ``bottom'' half remains unaltered during the course of the simulation, making $\hat{\mathbf{i}}$ an intrinsic measure of the RBC orientation that is not available from the inertia tensor.}
\end{figure}

\subsection{Numerical model}

A hybrid immersed-boundary-lattice-Boltzmann-finite-element model has been used \cite{kruger2011efficient, kruger2011particlestress}. The liquid phase is fully considered and solved by the lattice-Boltzmann method \cite{succi_lattice_2001, aidun2010lattice}, while the fluid-particle interaction is realised by the immersed-boundary method \cite{peskin_ibm_2002}. 
The red blood cells are represented by a finite-element mesh consisting of 1620 triangular facets. 
Physically, the cell membrane is made of a lipid bilayer (being essentially an incompressible 2D fluid), giving rise to a finite resistance against bending and changes of surface area, and a cytoskeleton, leading to a finite shear resistance \cite{mohandas_mechanical_1994,svetina_cooperative_2004,gompper2008soft}. 
The ensueing effects on the collective and rheological properties of the RBC suspension can be captured by assuming a total membrane energy of the form
\begin{equation}
 E = E_S + E_B + E_A + E_V\,,
\label{eq_energy_tot}
\end{equation}
where 
\begin{equation}
 E_S = \oint dA \left[ \frac{\kappa_S}{12} (I_1^2 + 2I_1 -2I_2) + \frac{\kappa_\alpha}{12}I_2^2 \right]
\label{eq_shear_energy}
\end{equation}
describes the energy penalty against shear and area dilation,
\begin{equation}
 E_B = \frac{\kappa_B}{2} \oint dA\left( H - H^{(0)}\right)^2 
\label{eq_bending_energy}
\end{equation}
is energy change due to bending, while
\begin{equation}
  E_A = \frac{\kappa_A}{2} \frac{(A-A^{(0)})^2}{A^{(0)}} 
\label{eq_area_energy}
\end{equation}
and
\begin{equation}
  E_V = \frac{\kappa_V}{2} \frac{(V-V^{(0)})^2}{V^{(0)}}
\label{eq_volume_energy}
\end{equation}
penalize changes of membrane surface area $A$ and cell volume $V$ over their undeformed counterparts, $A^{(0)}$ and $V^{(0)}$, respectively.
Eq.~\eqref{eq_shear_energy} represents Skalak's energy density \cite{skalak_strain_1973} with $\mods$ and $\moda$ being the shear and area resistance, respectively. $I_1$ and $I_2$ denote the in-plane strain invariants, which are related to the local membrane deformation tensor (see \citet{kruger2011efficient} for more details). 
Note that both the parameters $\moda$ and $\modA$ penalize changes in the membrane surface area (physically, $\moda$ is related the cytoskeleton, while $\modA$ arises from the incompressibility of the the liquid bilayer). 
The bending energy in Eq.~\eqref{eq_bending_energy} has the classical Helfrich form \cite{helfrich_elastic_1973}, of which a simplified version
\begin{equation}
 E_\text{B} = \frac{\sqrt 3 \modb}{2} \sum_{\langle i, j \rangle} \left(\theta_{ij} - \theta^{\text{eq}}_{ij}\right)^2
\label{eq_bending_energy_discr}
\end{equation}
is employed in the simulations\cite{gompper_random_1996}. Here, $\modb$ is the bending modulus and the sum runs over all pairs of neighbouring facets of the tessellated RBC surface with mutual equilibrium normal-to-normal angles $\theta^{\text{eq}}_{ij}$. We remark that, in the present case, the biconcave shape of the RBC is not a result of the minimization of the bending energy subject to the constraint of fixed volume\cite{gompper2008soft}, but rather specified as input to the simulation via the construction of the membrane mesh (and ensured by the presence of the $\theta^{\text{eq}}_{ij}$ in Eq.~\eqref{eq_bending_energy_discr}).

In our simulations, a fixed number (494, 635, 776, and 917) of neutrally buoyant RBCs with equilibrium shape of a biconcave disk of radius of $r = 9$ (all quantities in lattice units) and equal internal and external viscosities have been distributed throughout a fixed fluid volume ($L_x \times L_y \times L_z = 100 \times 100 \times 160$) resulting in volume fractions $\Ht = 35$, $45$, $55$, and $65\%$, respectively. To improve numerical stability and avoiding particle overlap, a repulsive force inspired by \citet{feng_immersed_2004} is introduced. It is zero for distances larger than one lattice constant and behaves like $1/r^2$ for smaller distances. The effect of the membrane energy, Eq.~\eqref{eq_energy_tot}, is realized in terms of local forces derived from the principle of virtual work\cite{kruger2011efficient}.

The shear flow (periodic in $x$- and $y$-directions) has been realised by moving the bottom and top walls at $z = 0$ and $z = 160$ in opposite directions along the $x$-axis with velocities $\pm \Uwall$, giving rise to an average shear rate of $\gammadot=2\Uwall/L_z$. Inertia is neglected. One layer of RBCs has been stuck to each wall to avoid slip. A typical simulation snapshot illustrating the geometry is shown in Fig.\ \ref{subfig_snapshot}.

The elasticity parameter values $\mods$ and $\modb$ in simulation units can be found by matching the relevant dimensionless parameters of the problem (in this case the capillary number $\eta_0 \dot{\gamma}r / \mods$ and the reduced bending modulus $\modb / (\mods r^2)$). Two different RBC rigidities have been considered. The softer RBCs, $(\mods, \modb, \moda, \modA, \modV) = (0.02, 0.004, 1, 1, 1)$, correspond to healthy RBCs ($\mods = 5 \cdot 10^{-6}\, \text{N}\, \text{m}^{-1}$ and $\kappa_\text{B} = 2 \cdot 10^{-19}\, \text{N}\, \text{m}$) \cite{gompper2008soft}. The more rigid RBCs obey $(\mods, \modb, \moda, \modA, \modV) = (0.06, 0.012, 1, 1, 1)$. The former are denoted `s', the latter `r'. The choice of $\moda=\modA=\modV=1$ ensures that local area and volume deviations are restricted to a few percent.

In order to improve the statistics, all simulations have been repeated with various random initial RBC configurations. Ten and five independent runs have been performed for the softer and more rigid RBCs, respectively. All reported observables are averaged over independent runs and time in the steady state. Due to the absence of thermal fluctuations in the present model, all observed effects are shear-induced.

\subsection{Characterisation of the flow}

Since we are interested in bulk properties rather than wall-induced effects, we first examine RBC concentration, suspension velocity, and shear stress profiles to see whether a bulk-like behavior may be expected.

Fig.\ \ref{subfig_profiles} reveals that, within the inner $50\%$ of the volume between the walls (henceforth called the central region), the volume fraction is nearly constant and the velocity is linear, defining a constant shear rate. The shear stress is confirmed to be constant throughout the entire volume as expected for simple shear flow \cite{kruger2011particlestress}. Thus, an effective bulk shear viscosity $\eta(\gammadot)  \equiv \sigma_{xz}(\gammadot) / \gammadot$ can be defined as the ratio of shear stress $\sigma_{xz}$ and shear rate $\gammadot$ using the data in the central region (Fig. \ref{fig_supp_viscosities}). 
For the present purposes, the viscosity is only needed to define an effective capillary number (see eq.~\eqref{eq_ca_eff} below), which will turn out to be the central quantity governing the cell dynamics. A detailed investigation of the rheology will be presented in another publication.

The results analysed in the central region are expected to be characteristic of the bulk properties of the system. In the following, system properties will thus be determined in the central region where all studied observables have been found to be rather independent of the transverse position $z$.

\section{Results and discussion}
\label{sec_results}

\subsection{Transition from tumbling to tank-treading-like dynamics}
\label{subsec_tbtt}

\begin{figure*}[t]
 \centering
 \subfigure[\label{subfig_omega_cabare} angular velocity]{\includegraphics[width=0.44\linewidth]{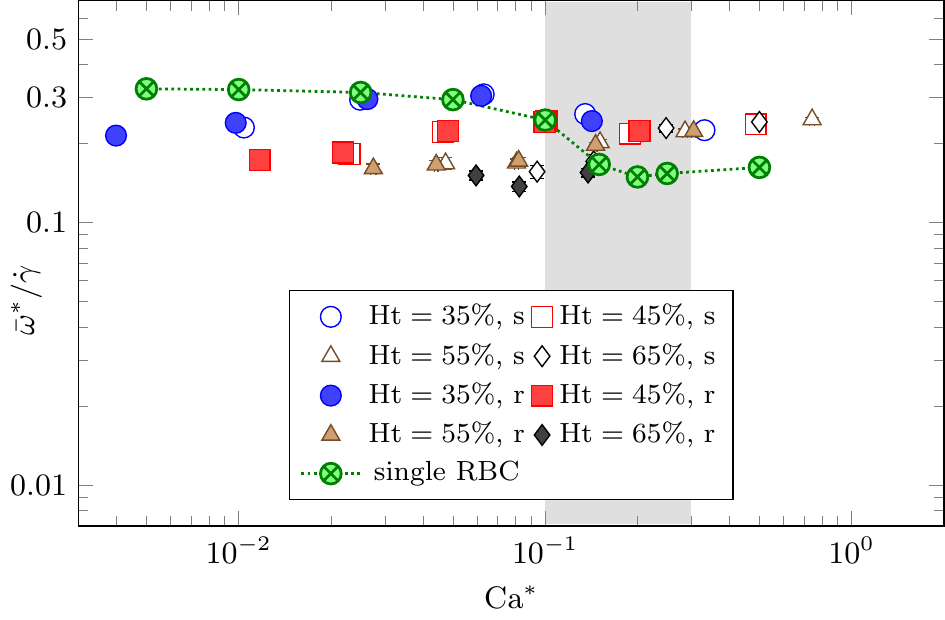}}\hfill
 \subfigure[\label{subfig_omega_caeff} tumbling frequency]{\includegraphics[width=0.44\linewidth]{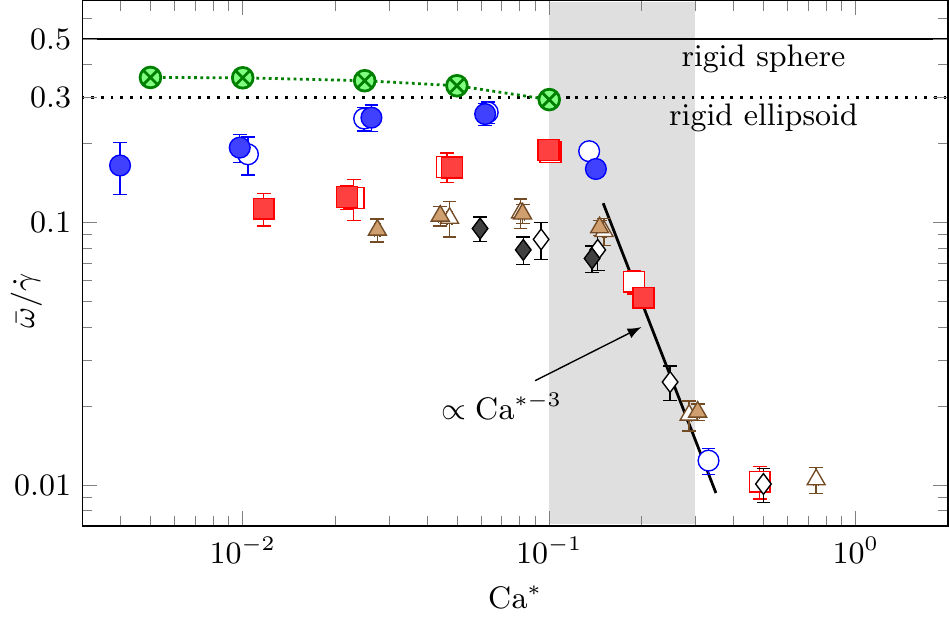}}
 \caption{\label{fig_rheology_omega_stress}  Reduced (a) angular velocity $\bar\omega^*/\gammadot$ and (b) tumbling frequency $\bar\omega/\dot\gamma$ versus effective capillary number $\Ca^*$. The legend applies to both panels. The grey area denotes the \TBTT. As seen in panel (b), for $\Ca^* > 0.1$, the average tumbling frequency of a single isolated cell is zero, whereas $\bar\omega/\dot\gamma$ approximately decays like  $\bar\omega/\gammadot\propto{\text{Ca}^*}^{-3}$ (inclined solid line) in the dense suspensions. Contrarily, the angular membrane velocity in (a) is rather independent of $\Ca^*$. Note that $\bar\omega^*/\gammadot$ and $\bar\omega/\gammadot$ are nearly identical for $\Ca^*\lesssim 0.1$. The analytic values of $\bar\omega/\dot\gamma$ for a rigid sphere and ellipsoid with aspect ratio $p=a/c=1.1/0.36$ are also shown in (b).}
\end{figure*}

\begin{figure}
 \centering
 \subfigure[$\dot \gamma\, t = 0$]{\includegraphics[width=0.32\linewidth]{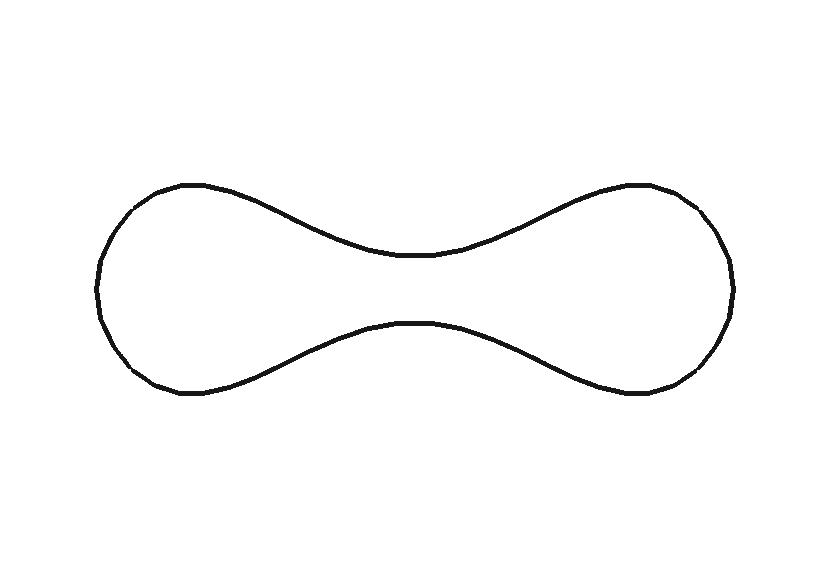}}
 \subfigure[$\dot \gamma\, t = 1.25$]{\includegraphics[width=0.32\linewidth]{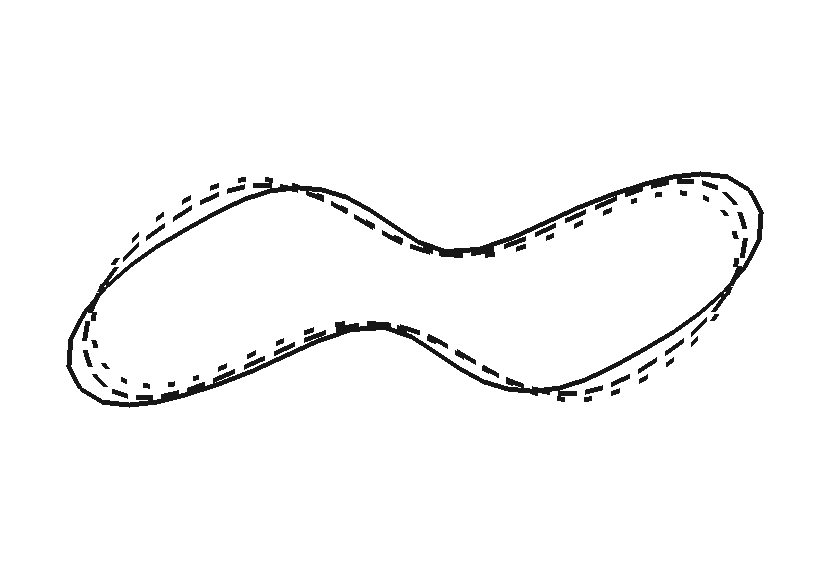}}
 \subfigure[$\dot \gamma\, t = 2.5$]{\includegraphics[width=0.32\linewidth]{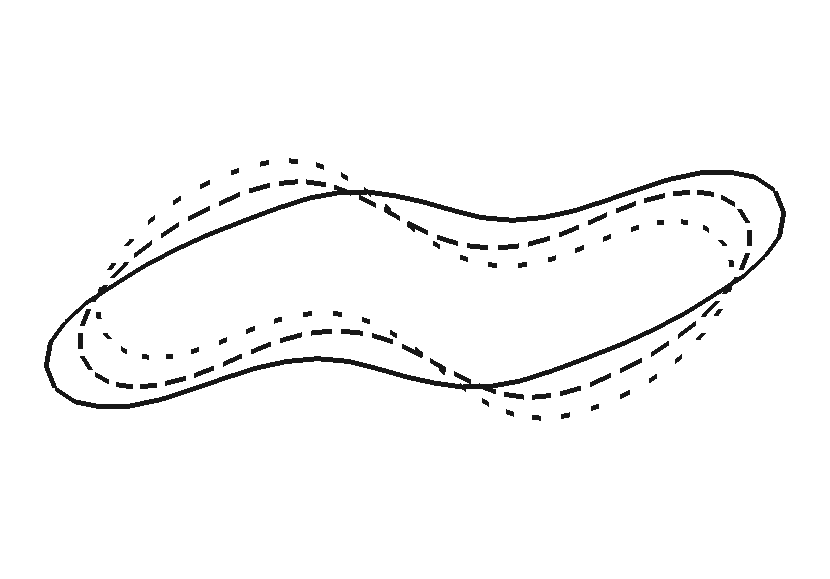}} \\
 \subfigure[$\dot \gamma\, t = 3.75$]{\includegraphics[width=0.32\linewidth]{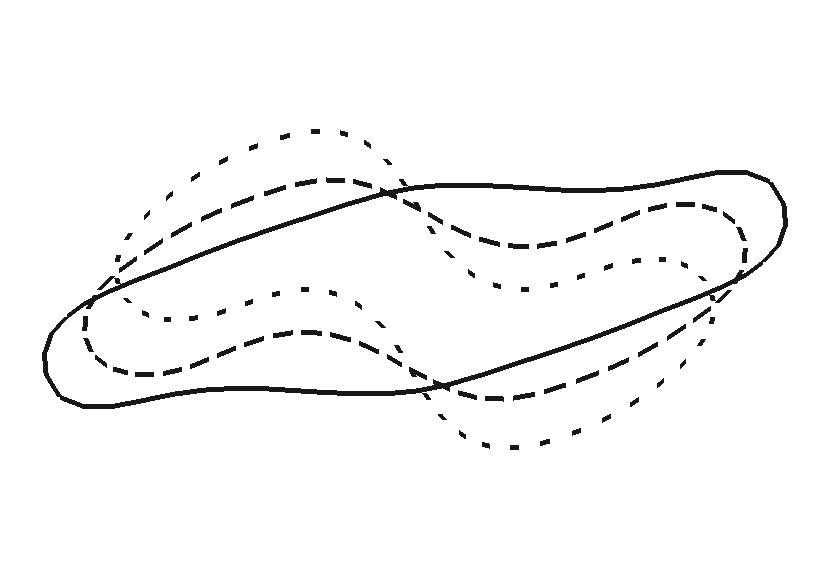}}
 \subfigure[$\dot \gamma\, t = 5$]{\includegraphics[width=0.32\linewidth]{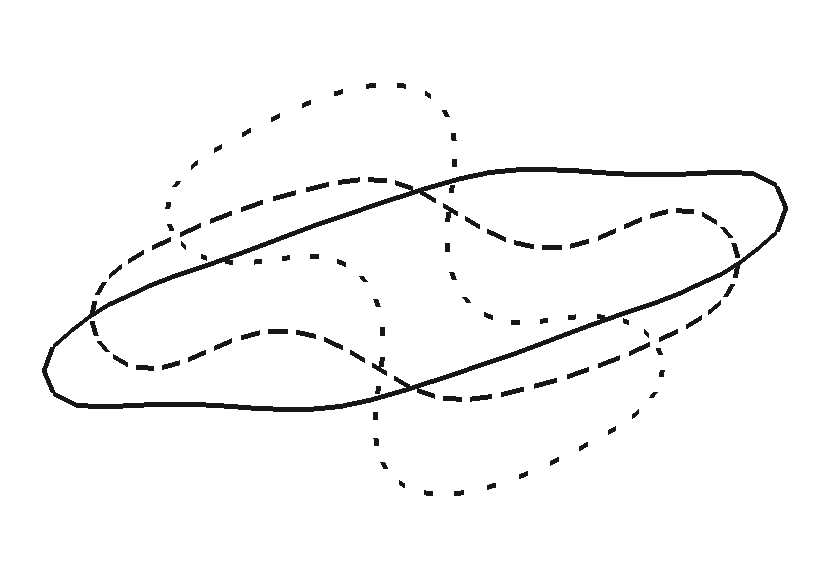}}
 \subfigure[$\dot \gamma\, t = 6.25$]{\includegraphics[width=0.32\linewidth]{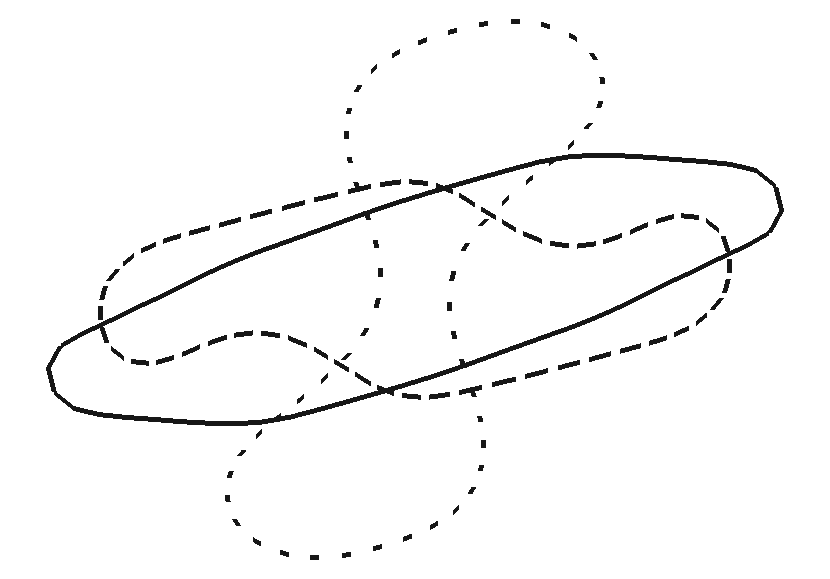}} \\
 \subfigure[$\dot \gamma\, t = 7.5$]{\includegraphics[width=0.32\linewidth]{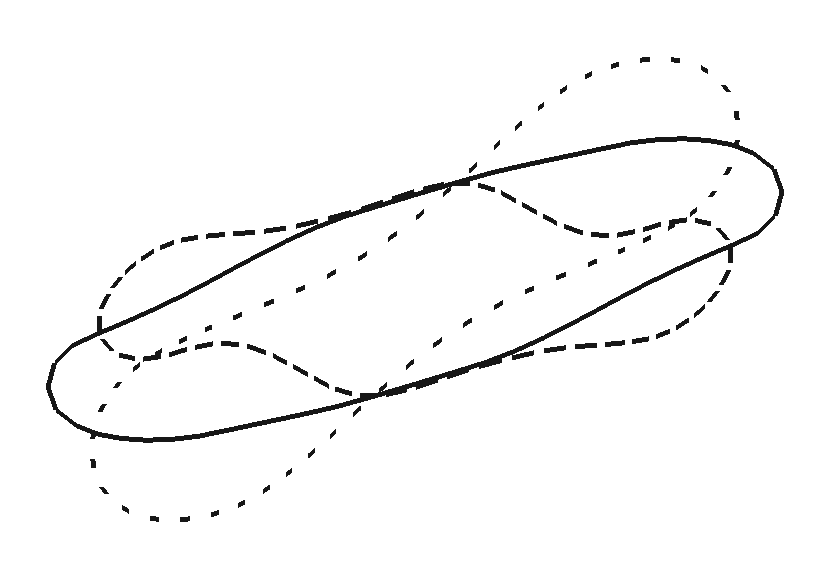}}
 \subfigure[$\dot \gamma\, t = 8.75$]{\includegraphics[width=0.32\linewidth]{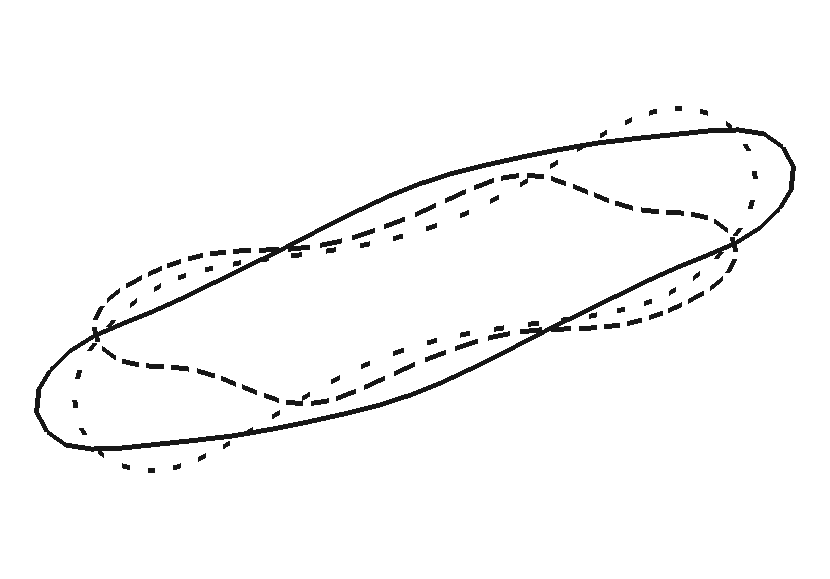}}
 \subfigure[$\dot \gamma\, t = 10$]{\includegraphics[width=0.32\linewidth]{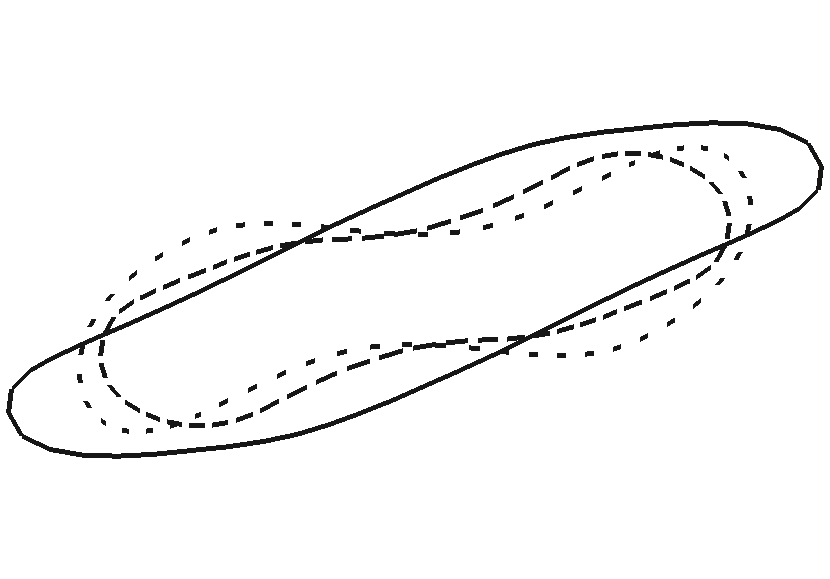}}
 \caption{\label{fig_rheology_TBTT} This sequence shows the time evolution of the rotational behavior of a single RBC in an external shear flow with rate $\dot \gamma$. The cross-section is parallel to the shearing plane, and the vorticity of the shear flow is clockwise. The evolution as function of dimensionless time $\dot \gamma t$ is shown for three different capillary numbers, $\Ca = 0.1$ (loosely dashed line), $0.2$ (densely dashed line), and $0.5$ (solid line). For the smallest value ($\Ca = 0.1$), the RBC performs a tumbling motion since it is not sufficiently deformed to undergo a tank-treading motion. However, already for $\text{Ca} = 0.2$, the RBC can rotate without tumbling, but shape oscillations are still observed. The tank-treading motion is fully developed for the highest capillary number ($\Ca = 0.5$). Note that in the dilute limit $\text{Ca}^* \approx \text{Ca}$ holds.}
\end{figure}

\begin{figure}
 \centering
 \subfigure[{} tumbling rotation]{\includegraphics[width=0.49\linewidth]{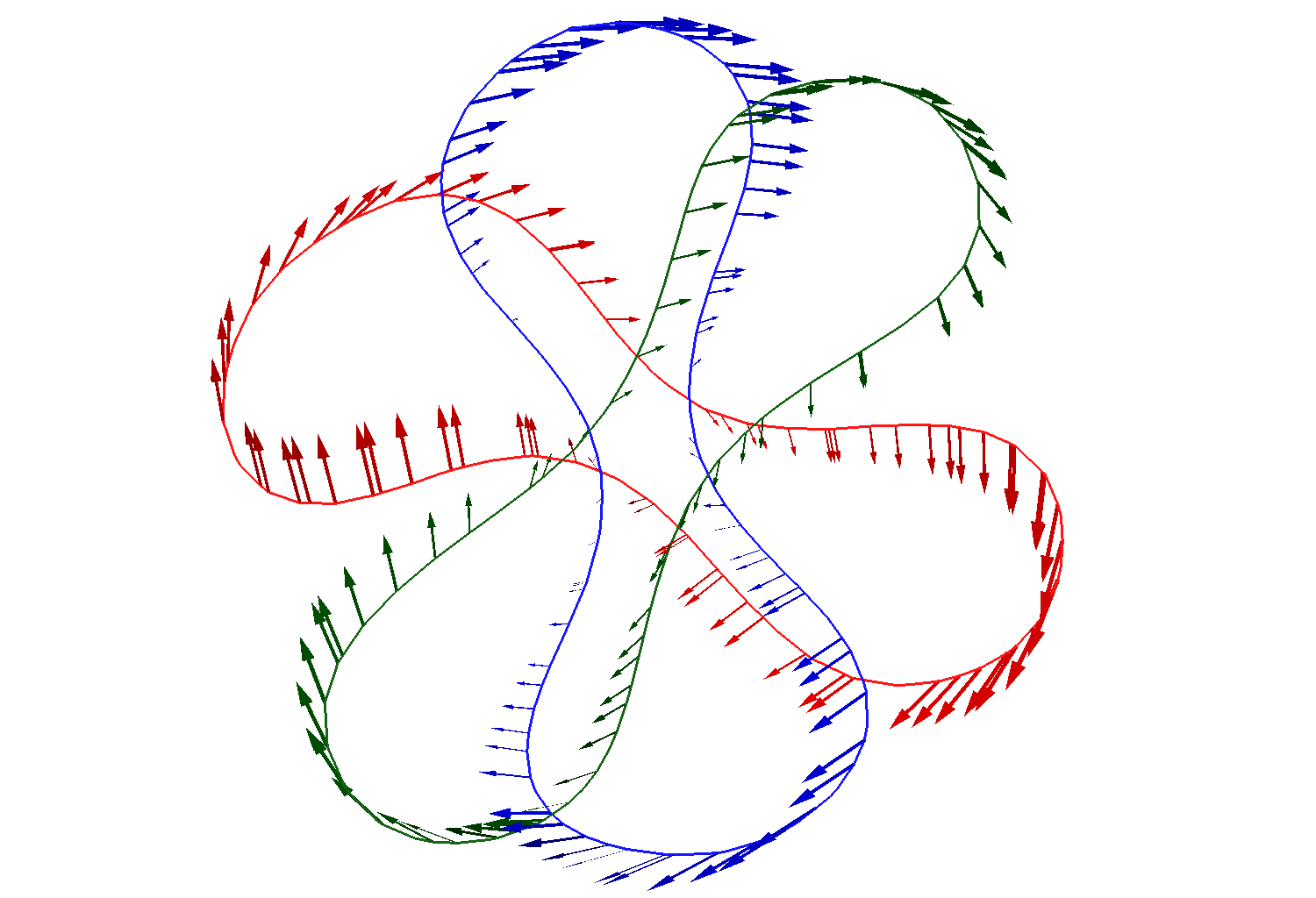}}
 \subfigure[{} tank-treading-like rotation]{\includegraphics[width=0.49\linewidth]{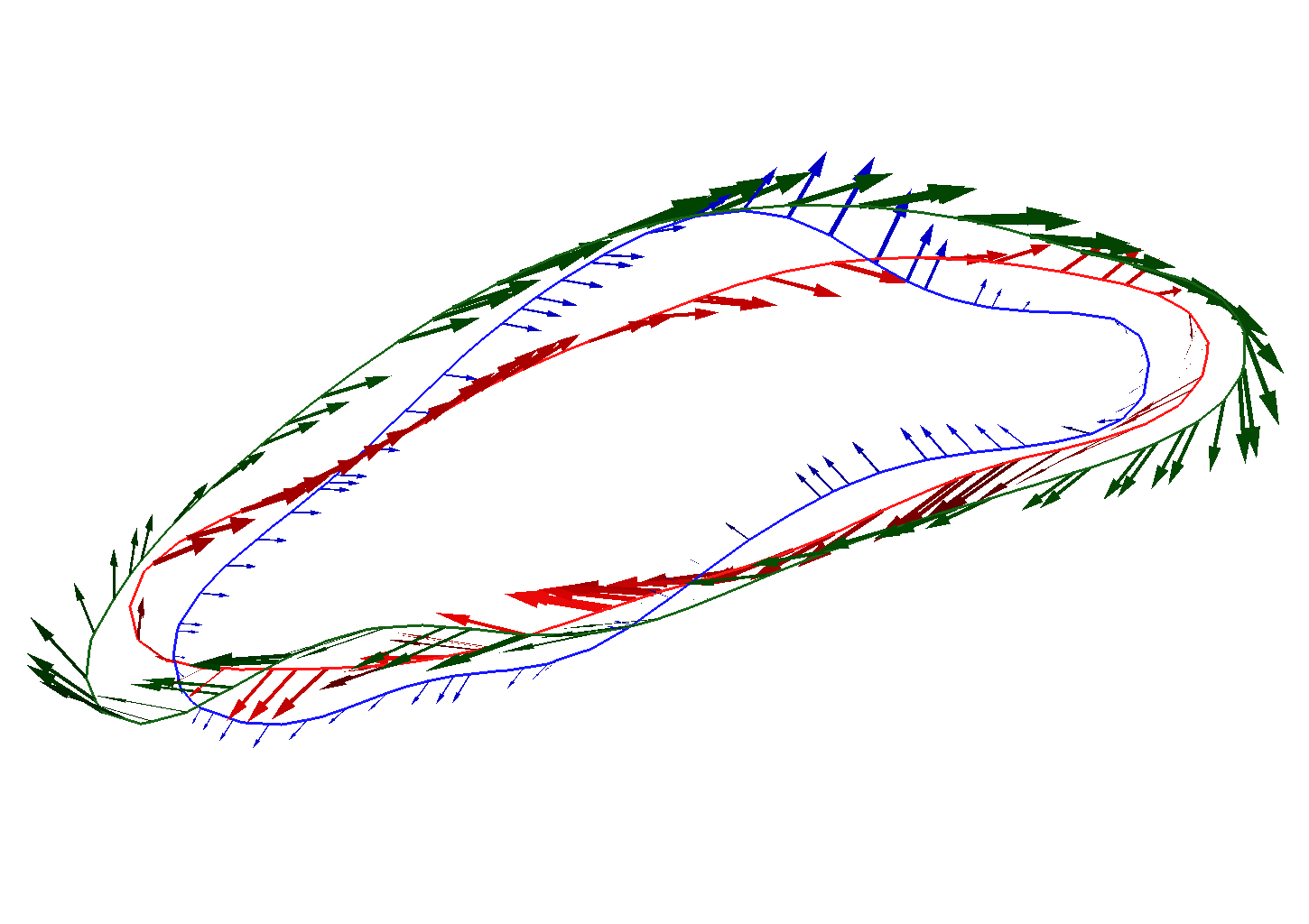}}
 \caption{\label{fig_supp_tank-treading-like} Exemplary cross-sections of a rotating RBC in a $45\%$-suspension at (a) $\Ca^* = 0.05$ and (b) $\Ca^* = 0.49$. The cross-sections correspond to the shearing plane and run through the cell centre. The shear flow is in clockwise direction. In the left panel, a flipping event (rigid body-like rotation) is shown (temporal order: red, blue, green; time difference between snapshots is $0.14 / \gammadot$). The right panel illustrates the irregular tank-treading-like rotation (temporal order: red, blue, green; time difference between snapshots is $0.35 / \gammadot$). Shape fluctuations due to collisions with neighbouring particles are clearly visible. Arrows indicate the local surface velocity.}
\end{figure}

\begin{figure}
 \centering
 {\includegraphics[width=0.9\linewidth]{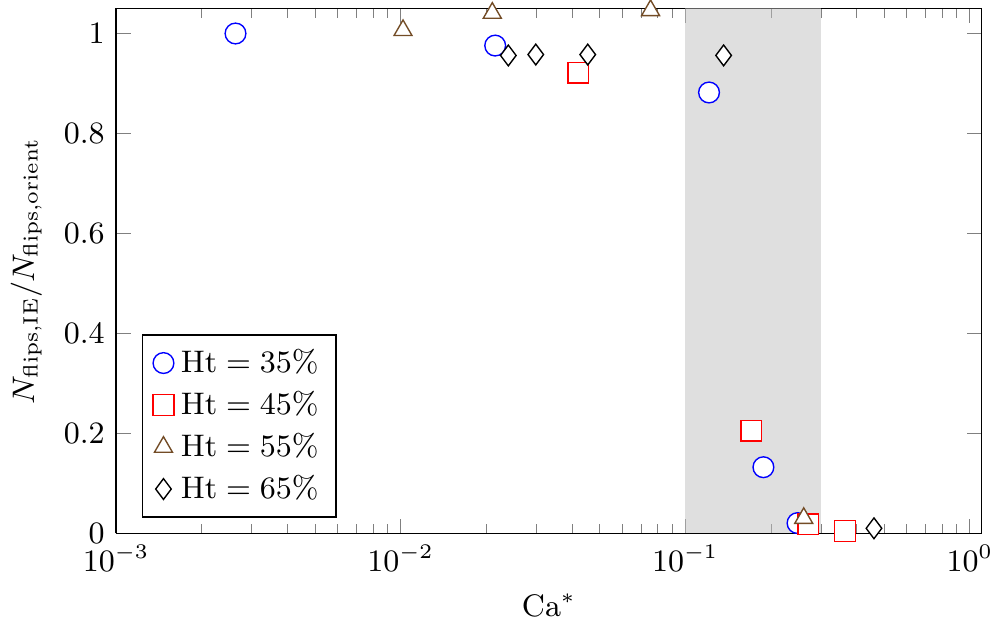}}
 \caption{\label{fig_fliprate}Ratio of the number of flips (i.e., half-turns) of the inertia ellipsoid, $N\st{flips,IE}$, and intrinsic cell orientation vector (see Fig.~\ref{fig_rbc_ellips}), $N\st{flips,orient}$, of RBCs in a suspension at different haematocrit values. At low effective capillary numbers, both inertia ellipsoid and orientation vector flip at the same rate (the flip number ratio being close to unity), characteristic for tumbling-like motion. In contrast, at high $\Ca^*$ only the intrinsic orientation vector performs flips, while the flip rate of the inertia ellipsoid is strongly reduced, indicating tank-treading-like motion.}
\end{figure}

We investigate the average RBC tumbling (rigid body-like) frequency $\bar \omega$ by tracking the orientation of the cells' inertia tensor in time. We recall the average tumbling frequency of a single rigid ellipsoid in a simple viscous shear flow,
\begin{equation}
 \frac{\bar\omega}{\gammadot} = \frac{1}{p+1/p},
\end{equation}
where $p=a/c$ ($a \geq b \geq c$ are the semi-axes of the ellipsoid, $a$ and $c$ are aligned with the shearing plane, see Fig.~\ref{fig_rbc_ellips}) \cite{jeffery1922motion}. For a rigid sphere ($p=1$), the tumbling frequency is $\omega/\dot\gamma=\frac{1}{2}$. One finds for the inertia ellipsoid of an undeformed RBC $a=b=1.1r$ and $c=0.36r$, and Jeffery's solution \cite{jeffery1922motion} predicts $\bar\omega/\dot\gamma=0.30$. On the contrary, for purely tank-treading cells, the average is $\bar\omega=0$, as they do not show any tumbling activity.

Additionally, we define the instantaneous reduced angular velocity of a RBC as
\begin{equation}
 \frac{\bar \omega^*}{\gammadot} := \frac{L_y}{ac \gammadot},
\end{equation}
where $ac$ is approximately the RBC cross-sectional area in the shearing plane and
\begin{equation}
 \mathbf L = \frac{1}{A} \oint \text{d}A\, (\mathbf r \times \mathbf v)
\end{equation}
is the surface-averaged rotational component of the membrane velocity ($\mathbf r$ and $\mathbf v$ are location and velocity of a RBC surface element relative to the current RBC centre, respectively, and $A$ is the RBC surface area). Note that $\bar \omega^*$ is sensitive to any form of membrane rotation, even if the cell's inertia tensor is not rotating in space (\emph{i.e.}, if there is no tumbling).

A first evidence for the onset of the \TBTT\ is provided in Fig.\ \ref{fig_rheology_omega_stress} where the reduced angular membrane velocity $\bar\omega^*/\gammadot$ and the reduced tumbling frequency $\bar\omega/\gammadot$ are plotted versus the effective capillary number
\begin{equation}
 \Ca^*= \frac{\eta \gammadot r}{\mods}
\label{eq_ca_eff}
\end{equation}
which gives the relative strength of the suspension stress $\sigma_{xz} = \eta \gammadot$, acting on the cells and the characteristic membrane stress, $\mods/r$. The approximately constant behavior of $\bar\omega^*$ in panel (a) indicates that rotational motion is always present in the studied $\Ca^*$-range. Despite this fact, we observe a rather sharp decrease of the tumbling frequency at $\Cacrit \simeq 0.2$ in panel (b). This clearly shows that, for $\Ca^* > \Cacrit$, a type of rotational motion is dominant which allows for a non-rotating tensor of inertia. Tank-treading-like dynamics provides such an alternative. 
Interestingly, for an isolated RBC we see a \TBTT in the same region (Fig.\ \ref{fig_rheology_TBTT}), which is in qualitative agreement with previous works \cite{pozrikidis_effect_2001, pozrikidis_numerical_2003, zhang_effects_2009}.
Furthermore, at $\sigma\approx 0.4\,\text{Pa}$, which corresponds to $\Ca^* \approx 0.2$ for healthy RBCs, the transition has been observed experimentally for single RBCs \cite{forsyth2011multiscale}. 
Note that all data for $\Ca^* > \Cacrit$ collapse onto a single master curve, which is not the case when plotted as function of the bare capillary number Ca (not shown). The decay of the tumbling frequency in the region $\text{Ca}^{*} \approx 0.2 \text{--} 0.3$ can be captured by a simple power law, $\bar\omega/\gammadot\propto{\text{Ca}^*}^{-3}$. 
We have to emphasize, however, that this fit is entirely ad-hoc. The establishment of such a power-law requires a more careful analysis which goes beyond the scope of the present work.

The picture of a \TBTT\ characterized by a critical capillary number without explicit dependence on haematocrit is further supported by a survey of the dynamics at the level of the individual cells. 
Fig.~\ref{fig_supp_tank-treading-like} exemplifies the typical temporal behaviour of an individual RBC in a moderately dense suspension, from which one can clearly identify tumbling motion at low effective capillary numbers, whereas tank-treading is performed at large $\Ca^*$. 
Interestingly, even at this comparatively large value of haematocrit (45\%), the typical rotational motion of an individual cell does not appear to be significantly different from that of a single cell, apart from occasional hydrodynamical collisions with its neighbours.
While a more detailed analysis of these aspects will be presented in a separate work, we point out here that a comparison of the rate of flips \cite{janoschek2011rotational} of the equivalent inertia ellipsoid on the one hand and of a suitable intrinsic membrane orientation vector\footnote{The intrinsic membrane orientation vector is computed at the beginning of the simulation as the vector connecting the centre of masses of the top and bottom halves of the undeformed RBC, see Fig.~\ref{fig_rbc_ellips}.} on the other hand provides an independent means to distinguish tumbling from tank-treading behaviour. 
This is demonstrated in Fig.~\ref{fig_fliprate}, where we plot the ratio of the total number of flips (obtained by integrating the flip rate over several decades in inverse shear rate) of the two orientation vectors.
Consistent with the expectations from the analysis of the angular velocities, we find that for low effective capillary numbers, both the inertia ellipsoid and intrinsic membrane orientation vector flip at the same rate, indicative of tumbling-like motion. In contrast, at higher $\Ca^*$, only the intrinsic membrane orientation vector performs flips while the orientation of the inertia ellipsoid remains virtually fixed, as would be expected for tank-treading-like motion.

It deserves special notice that, while the \TBTT\ is fairly well understood in terms of flow gradient effects and cell deformability (it is expected to occur when the fluid stress is strong enough to push the cells through the maximum of the elastic energy)\cite{skotheim2007red}, its occurrence in a dense suspension, where collective effects play a major role, is a more complex phenomenon. Our results suggest that, even at a haematocrit of $\Ht=65\%$, RBCs can perform a tank-treading-like motion in an effective medium whose viscosity is no longer the viscosity of the ambient fluid but the significantly higher effective suspension viscosity. It has also to be noted that, due to RBC collisions and related complicated dynamics, an unperturbed swinging or tank-treading motion for $\Ca\gtrsim\Cacrit$ is not expected. Instead, the cells' inclination and deformation fluctuate about their time averages, which can also be inferred from Fig.\ \ref{fig_supp_tank-treading-like}. Therefore, the term \emph{tank-treading-like} is used to emphasise that the cells are not tumbling, but neither show a perfect, unperturbed tank-treading motion.

\subsection{Cell alignment and orientational ordering}
\label{subsec_order}

\begin{figure*}
 \centering
 \subfigure[\label{subfig_rheology_inclination_1} order parameter]{\includegraphics[height=0.26\linewidth]{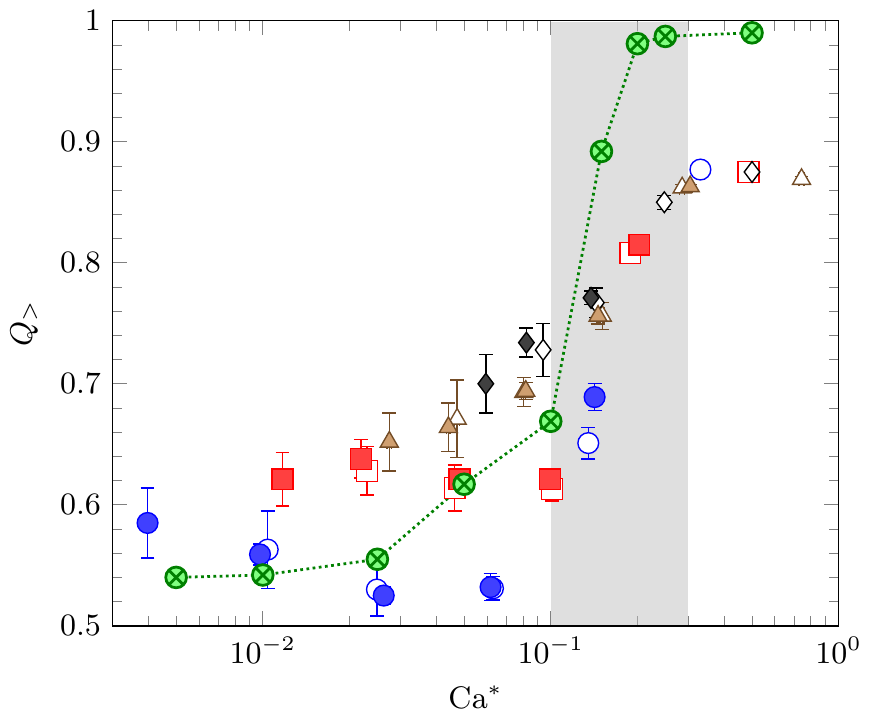}}
 \subfigure[\label{subfig_rheology_inclination_2} inclination angle]{\includegraphics[height=0.26\linewidth]{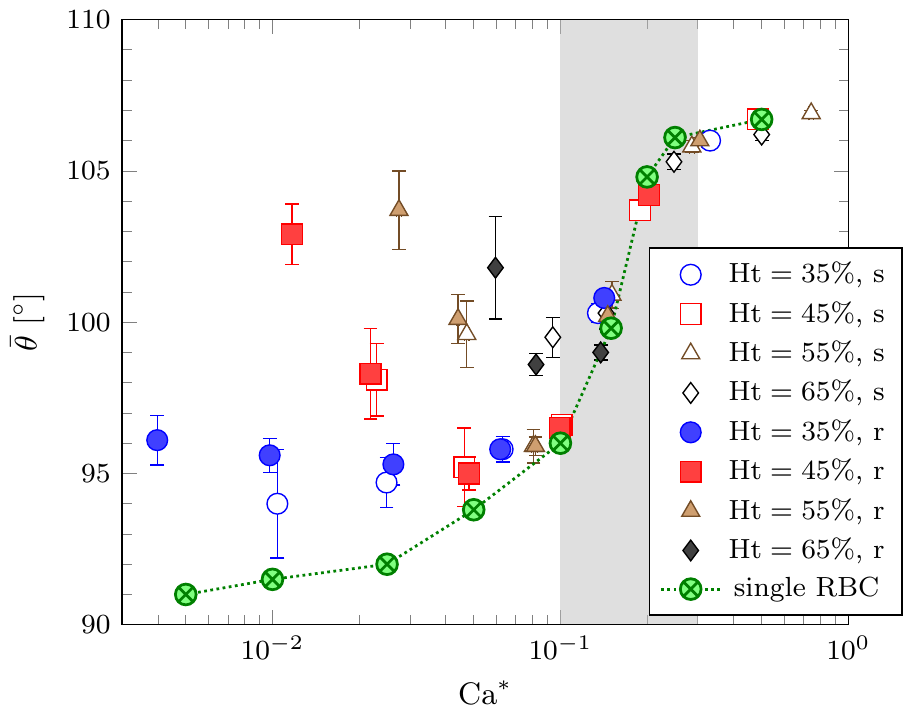}}
 \subfigure[\label{subfig_rheology_inclination_3} inclination angle probability]{\includegraphics[height=0.26\linewidth]{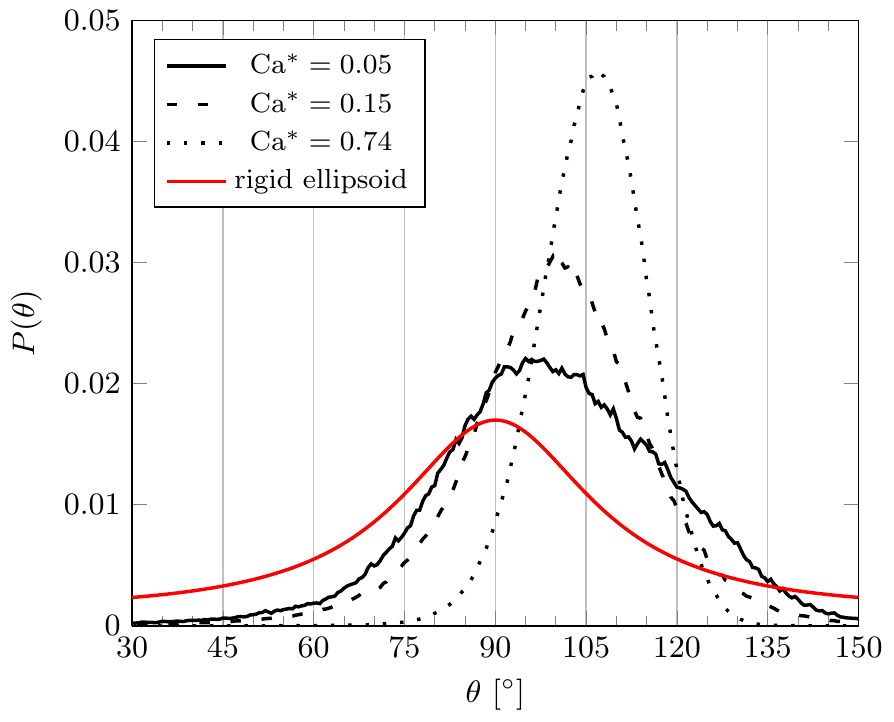}}
 \caption{\label{fig_rheology_inclination} (a) Order parameter and (b) average inclination angle versus effective capillary number $\Ca^*$. The curves obtained for one single cell are also shown. The grey area denotes the \TBTT. The legend applies to both panels. (c) Inclination angle probability distribution for the softer cells with $\Ht=55\%$ and selected values of $\Ca^*$. The analytic curve for a single rigid ellipsoid (aspect ratio $p = a/c = 1.1/0.36$) is also shown. The single cell in panel (b) approaches $90^\circ$ for decreasing $\Ca^*$, as expected from the analytical solution.}
\end{figure*}

\begin{figure*}
 \centering
 \subfigure[\label{subfig_deformation_1} deformation probability]{\includegraphics[width=0.48\linewidth]{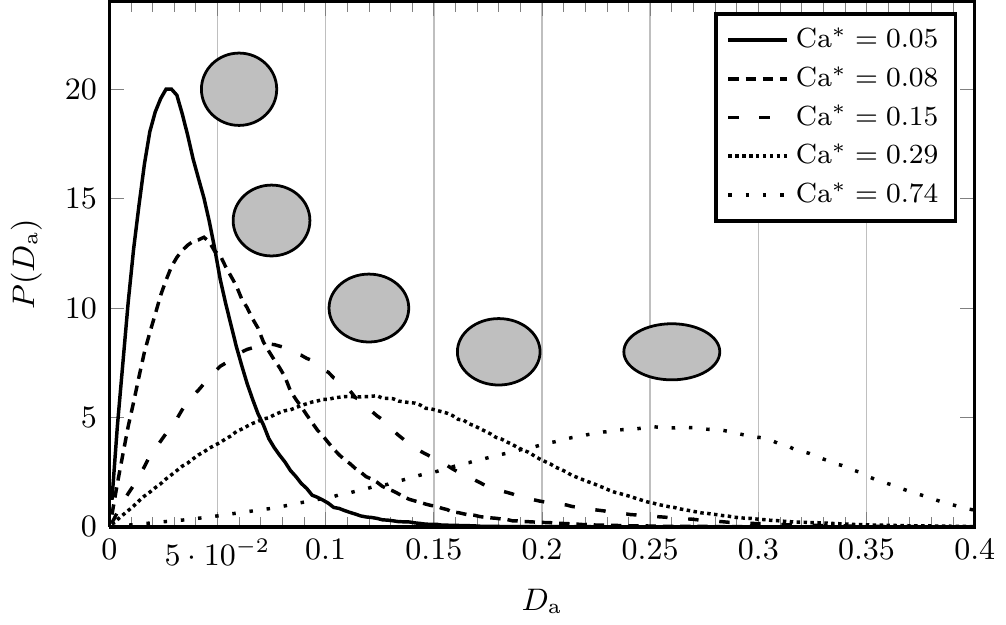}}\hfill
 \subfigure[\label{subfig_deformation_2} deformation with maximum probability]{\includegraphics[width=0.46\linewidth]{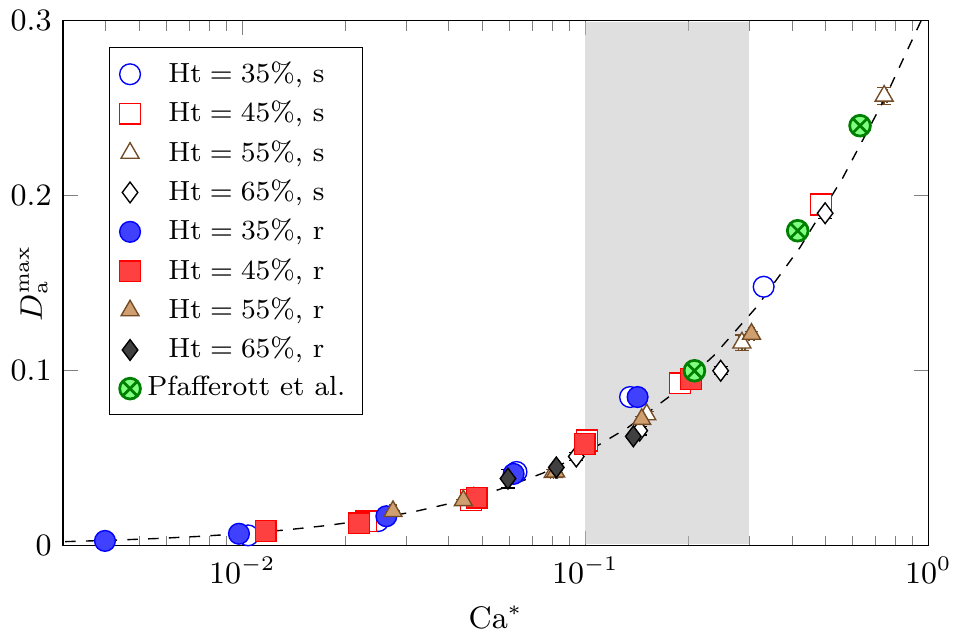}}
 \caption{\label{fig_deformation} (a) Deformation probability distribution for the softer cells at $\Ht = 55\%$ for different effective capillary numbers $\Ca^*$. For each curve, an ellipsoid is shown whose deformation parameter equals $D_{\text{a}}^{\text{max}}$, the deformation parameter with maximum probability. (b) $D_{\text{a}}^{\text{max}}$ versus $\Ca^*$. The power-law fit, $a/b|_{\text{max}} - 1 = 0.89 \times {\Ca^*}^{0.9}$, is shown as dashed line. There is a one-to-one correspondence between $a/b$ and $D_{\text{a}}$. The grey area denotes the \TBTT. Note the excellent agreement with the experimental data obtained for young RBCs reported in Fig.~2 in \citet{pfafferott1985red} (we converted the shear stress reported therein to $\Ca^*$ by assuming a shear elasticity of $\mods = 5\, \upmu\text{N}/\text{m}$ \cite{gompper2008soft}).}
\end{figure*}

The onset of the \TBTT\ seems to be accompanied by a transition in the orientational order parameter $Q_>$ of the system which is defined as the largest eigenvalue of the order tensor \cite{saupe1968recent}
\begin{equation}
 Q_{\alpha\beta} := \frac{1}{2} \langle 3 \hat o_{\alpha} \hat o_{\beta} - \delta_{\alpha \beta}\rangle, \quad \alpha,\beta\in\{x,y,z\}.
\end{equation}
The RBC orientation vector $\hat{\mathbf o}$ is defined as the inertia tensor eigenvector corresponding to the shortest semi-axis $c$ (see Fig.~\ref{fig_rbc_ellips}). The related eigenvector of $\mathbf Q$ is called the director $\mathbf n$ indicating the average orientation of the cells.

As illustrated in Fig.\ \ref{fig_rheology_inclination}, both $Q_>$ and the average inclination angle $\bar \theta$ (average angle between the director and the flow axis) change rapidly at $\Ca^*\sim\Cacrit$. We have also investigated the time evolution and spatial dependence of $\mathbf n$ and $Q_>$. Both quantities show only slight fluctuations about their averages. Note that RBC ordering at high shear rates has been observed experimentally \cite{schmid-schonbein_fluid_1969}. Similarly to the behavior of the tumbling frequency, all simulated data follow a master curve for $\Ca^*\ge\Cacrit$ and can be described by one single rather than by two independent parameters ($\text{Ca}^*$ instead of $\Ca$ and $\Ht$).

In qualitative agreement with analytical predictions \cite{abkarian2007swinging, skotheim2007red, vlahovska_dynamics_2011} and previous simulations \cite{bagchi_dynamics_2009, clausen_capsule_2010}, the average inclination angle approaches $90\deg$ in the limit of vanishing capillary number \cite{jeffery1922motion} and grows with increasing capillary number toward the onset of tank-treading (it is bounded from above by $135\deg$). 
In the tank-treading regime, the average inclination angle is expected to decrease again \cite{barthes-biesel_motion_1980, bagchi_dynamics_2009, vlahovska_dynamics_2011}, as the capsule becomes more elongated and aligns further with the flow. For the limited range of capillary numbers accessible to present simulations, however, we do not observe such a behavior. Rather, the inclination angle is found to remain roughly constant or even slightly increases (which is, however, in line with the predictions of \citet{skotheim2007red}).

A typical distribution of inclination angles is shown in Fig.\ \ref{subfig_rheology_inclination_3}. For comparison, the probability of finding the rigid inertia ellipsoid of an RBC (aspect ratio $p = a/c$) with a given orientation angle in shear flow is
\begin{equation}
 P(\theta) \propto \frac{p + \frac{1}{p}}{p\cos^2\theta + \frac{1}{p} \sin^2 \theta}.
\end{equation}
This relation can be extracted from the known expression for the angular velocity of the ellipsoid \cite{jeffery1922motion}.

As $\gammadot$ increases, the centre of $P(\theta)$ is shifted towards larger angles and its shape becomes narrower. While the former results in an increase of $\bar \theta$, the latter leads to a higher orientational order $Q_>$ since deviations from a given cell orientation become restricted to a narrower range. It is striking that, for $\Ca^* \gtrsim \Cacrit$, the average inclination angle equals that of the isolated cell (Fig.\ \ref{subfig_rheology_inclination_2}).

\subsection{Deformation behavior}
\label{subsec_deformation}

In Fig.\ \ref{subfig_deformation_1}, some typical deformation probability distributions $P(D_{\text{a}})$ are shown. Here,
\begin{equation}
 D_{\text{a}} := \frac{a/a_0 - b/b_0}{a/a_0 + b/b_0}
\end{equation}
is the Taylor deformation parameter as a measure for the RBC asymmetry in the $a$-$b$-plane, where the index zero refers to the undeformed state and $a_0 = b_0$ holds for an RBC. Using this data, we determine $D_{\text{a}}^{\text{max}}$, the deformation parameter with maximum probability. It is found (Fig.\ \ref{subfig_deformation_2}) that the data on $D_{\text{a}}^{\text{max}}$ collapse onto a single master curve when plotted as a function of $\Ca^*$ for all studied values of control parameters.

In marked contrast to the behavior of tumbling frequency and orientational order parameter, $D_{\text{a}}^{\text{max}}$ shows no signature of the \TBTT. A similar observation for dilute RBC suspensions has been reported before \cite{forsyth2011multiscale}. The entire observed range of capillary numbers can be parameterised by a simple power law, $a/b|_{\text{max}} - 1 \propto {\Ca^*}^{0.9}$, as shown in Fig.\ \ref{subfig_deformation_2}.

\section{Conclusions and outlook}
\label{sec_conclusions}

The focus of the present study is on the tumbling-to-tank-treading transition (\TBTT) in suspensions of aggregation-free red blood cells in the limit of high volume fractions where collective effects become dominant. We provide evidence that this transition occurs when the ratio of the effective suspension stress to the characteristic membrane stress (effective capillary number) exceeds a threshold value $\Cacrit$. Note that all dependence on volume fraction is implicit in $\Ca^*$ through its dependence on the effective stress. 
For a single cell, $\Cacrit$ corresponds to the stress where the cell is driven through the maximum of its elastic energy, thus allowing tank-treading-like dynamics.

The average tumbling frequency and cell inclination angle signal the onset of the \TBTT\ and a scaling collapse above $\Cacrit$ when expressed in dependence of the effective capillary number.
Remarkably, the most probable cell deformation changes continuously in the investigated parameter range following a scaling law, $\propto {\Ca^*}^{0.9}$, without any signature of the observed transition.
The value of $\Cacrit$ coincides with that for a transition from a less ordered cell orientation distribution to a highly ordered phase. 

The above findings support the following conclusions: (i) The cell dynamics is dominated by $\Ca^*$ and therefore the suspension stress. The cells in the tank-treading-like state behave more like isolated particles experiencing their environment only via the suspension stress. 
(ii) The large degree of orientational ordering at $\Ca^* \geq \Cacrit$ is related to the onset of tank-treading-like dynamics. (iii) Cell deformation alone does not seem to be a relevant factor for the \TBTT.

Interestingly, a scaling of static and dynamic quantities in terms of an effective, concentration-corrected dimensionless number is not uncommon in soft matter systems and has been observed previously, for instance, for polymer solutions \cite{hur_dynamics_2001,huang_semidilute_2010, singh_conformational_2012}.
An important difference between these and the present system is, however, the volume and area conservation of the RBCs, which implies that excluded volume effects should play a much more dominant role for RBC suspensions. This might be one of the reasons why many quantities lack scaling with $\Ca^*$ in the tumbling regime, where, in constrast to the tank-treading case, mutual disturbance of the cells is expected to be important.

During tank-treading, particles explore a smaller volume as compared to tumbling, thus providing a less strong hindrance to the motion of other particles. 
This is expected to affect other quantities, such as stresses and diffusivity, which are sensitive to the interactions between particles.
Indeed, it is commonly expected that the suspension viscosity is reduced as the number of tank-treading particles increases \cite{schmid-schonbein_fluid_1969, chien_shear_1970, forsyth2011multiscale, fedosov_predicting_2011}. 

Furthermore, our work opens a route to equipping coarse-grained blood models \cite{janoschek2010simplified, melchionna_model_2011} with proper constitutive laws for the tumbling/tank-treading probability or the average particle deformation given an ambient stress value.

Finally, while we focused on the limit of high $\Ca^*$, many interesting open questions remain regarding the opposite limit of small shear rates such as the possible existence of a yield stress or flow heterogeneity \cite{varnik2003shear, mandal2012heterogeneous}.

\section*{Acknowledgements}

This work is financially supported by the DFG-project VA205/5-1. We thank B.~Kaoui, S.~Frijters, J.~Harting, and M.~Ripoll for valuable discussions. We are also grateful for the computational time granted by the Juelich Supercomputing Centre (Project ESMI17). ICAMS acknowledges funding from its industrial sponsors, the state of North-Rhine Westphalia and the European Commission in the framework of the European Regional Development Fund (ERDF).

\providecommand*{\mcitethebibliography}{\thebibliography}
\csname @ifundefined\endcsname{endmcitethebibliography}
{\let\endmcitethebibliography\endthebibliography}{}


\begin{mcitethebibliography}{46}
\providecommand*{\natexlab}[1]{#1}
\providecommand*{\mciteSetBstSublistMode}[1]{}
\providecommand*{\mciteSetBstMaxWidthForm}[2]{}
\providecommand*{\mciteBstWouldAddEndPuncttrue}
  {\def\EndOfBibitem{\unskip.}}
\providecommand*{\mciteBstWouldAddEndPunctfalse}
  {\let\EndOfBibitem\relax}
\providecommand*{\mciteSetBstMidEndSepPunct}[3]{}
\providecommand*{\mciteSetBstSublistLabelBeginEnd}[3]{}
\providecommand*{\EndOfBibitem}{}
\mciteSetBstSublistMode{f}
\mciteSetBstMaxWidthForm{subitem}
{(\emph{\alph{mcitesubitemcount}})}
\mciteSetBstSublistLabelBeginEnd{\mcitemaxwidthsubitemform\space}
{\relax}{\relax}

\bibitem[Schmid-Sch{\"o}nbein and Wells(1969)]{schmid-schonbein_fluid_1969}
H.~Schmid-Sch{\"o}nbein and R.~Wells, \emph{Science}, 1969, \textbf{165},
  288--291\relax
\mciteBstWouldAddEndPuncttrue
\mciteSetBstMidEndSepPunct{\mcitedefaultmidpunct}
{\mcitedefaultendpunct}{\mcitedefaultseppunct}\relax
\EndOfBibitem
\bibitem[Skotheim and Secomb(2007)]{skotheim2007red}
J.~Skotheim and T.~Secomb, \emph{Phys. Rev. Lett.}, 2007, \textbf{98},
  78301\relax
\mciteBstWouldAddEndPuncttrue
\mciteSetBstMidEndSepPunct{\mcitedefaultmidpunct}
{\mcitedefaultendpunct}{\mcitedefaultseppunct}\relax
\EndOfBibitem
\bibitem[Abkarian \emph{et~al.}(2007)Abkarian, Faivre, and
  Viallat]{abkarian2007swinging}
M.~Abkarian, M.~Faivre and A.~Viallat, \emph{Phys. Rev. Lett.}, 2007,
  \textbf{98}, 188302\relax
\mciteBstWouldAddEndPuncttrue
\mciteSetBstMidEndSepPunct{\mcitedefaultmidpunct}
{\mcitedefaultendpunct}{\mcitedefaultseppunct}\relax
\EndOfBibitem
\bibitem[Sui \emph{et~al.}(2008)Sui, Chew, Roy, Cheng, and
  Low]{sui_dynamic_2008}
Y.~Sui, Y.~Chew, P.~Roy, Y.~Cheng and H.~Low, \emph{Phys. Fluids}, 2008,
  \textbf{20}, 112106\relax
\mciteBstWouldAddEndPuncttrue
\mciteSetBstMidEndSepPunct{\mcitedefaultmidpunct}
{\mcitedefaultendpunct}{\mcitedefaultseppunct}\relax
\EndOfBibitem
\bibitem[Keller and Skalak(1982)]{keller_motion_1982}
S.~Keller and R.~Skalak, \emph{J. Fluid Mech.}, 1982, \textbf{120},
  27--47\relax
\mciteBstWouldAddEndPuncttrue
\mciteSetBstMidEndSepPunct{\mcitedefaultmidpunct}
{\mcitedefaultendpunct}{\mcitedefaultseppunct}\relax
\EndOfBibitem
\bibitem[Beaucourt \emph{et~al.}(2004)Beaucourt, Rioual, S{\'e}on, Biben, and
  Misbah]{beaucourt2004steady}
J.~Beaucourt, F.~Rioual, T.~S{\'e}on, T.~Biben and C.~Misbah, \emph{Phys. Rev.
  E}, 2004, \textbf{69}, 011906\relax
\mciteBstWouldAddEndPuncttrue
\mciteSetBstMidEndSepPunct{\mcitedefaultmidpunct}
{\mcitedefaultendpunct}{\mcitedefaultseppunct}\relax
\EndOfBibitem
\bibitem[Noguchi and Gompper(2004)]{noguchi_fluid_2004}
H.~Noguchi and G.~Gompper, \emph{Phys. Rev. Lett.}, 2004, \textbf{93},
  258102\relax
\mciteBstWouldAddEndPuncttrue
\mciteSetBstMidEndSepPunct{\mcitedefaultmidpunct}
{\mcitedefaultendpunct}{\mcitedefaultseppunct}\relax
\EndOfBibitem
\bibitem[Kantsler and Steinberg(2006)]{kantsler2006transition}
V.~Kantsler and V.~Steinberg, \emph{Phys. Rev. Lett.}, 2006, \textbf{96},
  036001\relax
\mciteBstWouldAddEndPuncttrue
\mciteSetBstMidEndSepPunct{\mcitedefaultmidpunct}
{\mcitedefaultendpunct}{\mcitedefaultseppunct}\relax
\EndOfBibitem
\bibitem[Misbah(2006)]{misbah2006vacillating}
C.~Misbah, \emph{Phys. Rev. Lett.}, 2006, \textbf{96}, 028104\relax
\mciteBstWouldAddEndPuncttrue
\mciteSetBstMidEndSepPunct{\mcitedefaultmidpunct}
{\mcitedefaultendpunct}{\mcitedefaultseppunct}\relax
\EndOfBibitem
\bibitem[Forsyth \emph{et~al.}(2011)Forsyth, Wan, Owrutsky, Abkarian, and
  Stone]{forsyth2011multiscale}
A.~Forsyth, J.~Wan, P.~Owrutsky, M.~Abkarian and H.~Stone, \emph{Proc. Natl.
  Acad. Sci. USA}, 2011, \textbf{108}, 10986\relax
\mciteBstWouldAddEndPuncttrue
\mciteSetBstMidEndSepPunct{\mcitedefaultmidpunct}
{\mcitedefaultendpunct}{\mcitedefaultseppunct}\relax
\EndOfBibitem
\bibitem[Yazdani \emph{et~al.}(2011)Yazdani, Kalluri, and
  Bagchi]{yazdani2011tank}
A.~Yazdani, R.~Kalluri and P.~Bagchi, \emph{Phys. Rev. E}, 2011, \textbf{83},
  046305\relax
\mciteBstWouldAddEndPuncttrue
\mciteSetBstMidEndSepPunct{\mcitedefaultmidpunct}
{\mcitedefaultendpunct}{\mcitedefaultseppunct}\relax
\EndOfBibitem
\bibitem[Doddi and Bagchi(2009)]{doddi_three-dimensional_2009}
S.~Doddi and P.~Bagchi, \emph{Phys. Rev. E}, 2009, \textbf{79}, 046318\relax
\mciteBstWouldAddEndPuncttrue
\mciteSetBstMidEndSepPunct{\mcitedefaultmidpunct}
{\mcitedefaultendpunct}{\mcitedefaultseppunct}\relax
\EndOfBibitem
\bibitem[Clausen \emph{et~al.}(2011)Clausen, Reasor, and
  Aidun]{clausen_aidun_jfm2011}
J.~R. Clausen, D.~A. Reasor and C.~K. Aidun, \emph{J. Fluid. Mech.}, 2011,
  \textbf{685}, 1--33\relax
\mciteBstWouldAddEndPuncttrue
\mciteSetBstMidEndSepPunct{\mcitedefaultmidpunct}
{\mcitedefaultendpunct}{\mcitedefaultseppunct}\relax
\EndOfBibitem
\bibitem[Fedosov \emph{et~al.}(2011)Fedosov, Pan, Caswell, Gompper, and
  Karniadakis]{fedosov_predicting_2011}
D.~A. Fedosov, W.~Pan, B.~Caswell, G.~Gompper and G.~E. Karniadakis,
  \emph{Proc. Natl. Acad. Sci. USA}, 2011, \textbf{108}, 11772--11777\relax
\mciteBstWouldAddEndPuncttrue
\mciteSetBstMidEndSepPunct{\mcitedefaultmidpunct}
{\mcitedefaultendpunct}{\mcitedefaultseppunct}\relax
\EndOfBibitem
\bibitem[Reasor \emph{et~al.}(2013)Reasor, Clausen, and
  Aidun]{reasor_aidun_jfm2013}
D.~A. Reasor, J.~R. Clausen and C.~K. Aidun, \emph{J. Fluid. Mech.}, 2013,
  \textbf{726}, 497\relax
\mciteBstWouldAddEndPuncttrue
\mciteSetBstMidEndSepPunct{\mcitedefaultmidpunct}
{\mcitedefaultendpunct}{\mcitedefaultseppunct}\relax
\EndOfBibitem
\bibitem[Kr{\"u}ger \emph{et~al.}(2011)Kr{\"u}ger, Varnik, and
  Raabe]{kruger2011efficient}
T.~Kr{\"u}ger, F.~Varnik and D.~Raabe, \emph{Comput. Math. Appl.}, 2011,
  \textbf{61}, 3485--3505\relax
\mciteBstWouldAddEndPuncttrue
\mciteSetBstMidEndSepPunct{\mcitedefaultmidpunct}
{\mcitedefaultendpunct}{\mcitedefaultseppunct}\relax
\EndOfBibitem
\bibitem[Kr{\"u}ger \emph{et~al.}(2011)Kr{\"u}ger, Varnik, and
  Raabe]{kruger2011particlestress}
T.~Kr{\"u}ger, F.~Varnik and D.~Raabe, \emph{Phil. Trans. Roy. Soc. A}, 2011,
  \textbf{369}, 2414--2421\relax
\mciteBstWouldAddEndPuncttrue
\mciteSetBstMidEndSepPunct{\mcitedefaultmidpunct}
{\mcitedefaultendpunct}{\mcitedefaultseppunct}\relax
\EndOfBibitem
\bibitem[Succi(2001)]{succi_lattice_2001}
S.~Succi, \emph{{The Lattice Boltzmann Equation for Fluid Dynamics and
  Beyond}}, Oxford University Press, 2001, p. 368\relax
\mciteBstWouldAddEndPuncttrue
\mciteSetBstMidEndSepPunct{\mcitedefaultmidpunct}
{\mcitedefaultendpunct}{\mcitedefaultseppunct}\relax
\EndOfBibitem
\bibitem[Aidun and Clausen(2010)]{aidun2010lattice}
C.~Aidun and J.~Clausen, \emph{Annu. Rev. Fluid Mech.}, 2010, \textbf{42},
  439--472\relax
\mciteBstWouldAddEndPuncttrue
\mciteSetBstMidEndSepPunct{\mcitedefaultmidpunct}
{\mcitedefaultendpunct}{\mcitedefaultseppunct}\relax
\EndOfBibitem
\bibitem[Peskin(2002)]{peskin_ibm_2002}
C.~Peskin, \emph{Acta Numerica}, 2002, \textbf{11}, 479--517\relax
\mciteBstWouldAddEndPuncttrue
\mciteSetBstMidEndSepPunct{\mcitedefaultmidpunct}
{\mcitedefaultendpunct}{\mcitedefaultseppunct}\relax
\EndOfBibitem
\bibitem[Mohandas and Evans(1994)]{mohandas_mechanical_1994}
N.~Mohandas and E.~Evans, \emph{Annu. Rev. Bioph. Biom.}, 1994, \textbf{23},
  787--818\relax
\mciteBstWouldAddEndPuncttrue
\mciteSetBstMidEndSepPunct{\mcitedefaultmidpunct}
{\mcitedefaultendpunct}{\mcitedefaultseppunct}\relax
\EndOfBibitem
\bibitem[Svetina \emph{et~al.}(2004)Svetina, Kuzman, Waugh, Ziherl, and
  Zeks]{svetina_cooperative_2004}
S.~Svetina, D.~Kuzman, R.~Waugh, P.~Ziherl and B.~Zeks, \emph{Bioelectroch.},
  2004, \textbf{62}, 107--113\relax
\mciteBstWouldAddEndPuncttrue
\mciteSetBstMidEndSepPunct{\mcitedefaultmidpunct}
{\mcitedefaultendpunct}{\mcitedefaultseppunct}\relax
\EndOfBibitem
\bibitem[Gompper and Schick(2008)]{gompper2008soft}
G.~Gompper and M.~Schick, \emph{{Soft Matter: Lipid Bilayers and Red Blood
  Cells}}, Wiley-VCH, 2008\relax
\mciteBstWouldAddEndPuncttrue
\mciteSetBstMidEndSepPunct{\mcitedefaultmidpunct}
{\mcitedefaultendpunct}{\mcitedefaultseppunct}\relax
\EndOfBibitem
\bibitem[Skalak \emph{et~al.}(1973)Skalak, Tozeren, Zarda, and
  Chien]{skalak_strain_1973}
R.~Skalak, A.~Tozeren, R.~Zarda and S.~Chien, \emph{Biophys. J.}, 1973,
  \textbf{13}, 245--264\relax
\mciteBstWouldAddEndPuncttrue
\mciteSetBstMidEndSepPunct{\mcitedefaultmidpunct}
{\mcitedefaultendpunct}{\mcitedefaultseppunct}\relax
\EndOfBibitem
\bibitem[Helfrich(1973)]{helfrich_elastic_1973}
W.~Helfrich, \emph{Z. Naturforsch. C}, 1973, \textbf{28}, 693--703\relax
\mciteBstWouldAddEndPuncttrue
\mciteSetBstMidEndSepPunct{\mcitedefaultmidpunct}
{\mcitedefaultendpunct}{\mcitedefaultseppunct}\relax
\EndOfBibitem
\bibitem[Gompper and Kroll(1996)]{gompper_random_1996}
G.~Gompper and D.~Kroll, \emph{J. Phys. I}, 1996, \textbf{6}, 1305--1320\relax
\mciteBstWouldAddEndPuncttrue
\mciteSetBstMidEndSepPunct{\mcitedefaultmidpunct}
{\mcitedefaultendpunct}{\mcitedefaultseppunct}\relax
\EndOfBibitem
\bibitem[Feng and Michaelides(2004)]{feng_immersed_2004}
Z.-G. Feng and E.~Michaelides, \emph{J. Comput. Phys.}, 2004, \textbf{195},
  602--628\relax
\mciteBstWouldAddEndPuncttrue
\mciteSetBstMidEndSepPunct{\mcitedefaultmidpunct}
{\mcitedefaultendpunct}{\mcitedefaultseppunct}\relax
\EndOfBibitem
\bibitem[Jeffery(1922)]{jeffery1922motion}
G.~Jeffery, \emph{Proc. Roy. Soc. Lond. A Mat.}, 1922, \textbf{102},
  161--179\relax
\mciteBstWouldAddEndPuncttrue
\mciteSetBstMidEndSepPunct{\mcitedefaultmidpunct}
{\mcitedefaultendpunct}{\mcitedefaultseppunct}\relax
\EndOfBibitem
\bibitem[Pozrikidis(2001)]{pozrikidis_effect_2001}
C.~Pozrikidis, \emph{J. Fluid Mech.}, 2001, \textbf{440}, 269--291\relax
\mciteBstWouldAddEndPuncttrue
\mciteSetBstMidEndSepPunct{\mcitedefaultmidpunct}
{\mcitedefaultendpunct}{\mcitedefaultseppunct}\relax
\EndOfBibitem
\bibitem[Pozrikidis(2003)]{pozrikidis_numerical_2003}
C.~Pozrikidis, \emph{Ann. Biomed. Eng.}, 2003, \textbf{31}, 1194--1205\relax
\mciteBstWouldAddEndPuncttrue
\mciteSetBstMidEndSepPunct{\mcitedefaultmidpunct}
{\mcitedefaultendpunct}{\mcitedefaultseppunct}\relax
\EndOfBibitem
\bibitem[Zhang \emph{et~al.}(2009)Zhang, Johnson, and
  Popel]{zhang_effects_2009}
J.~Zhang, P.~Johnson and A.~Popel, \emph{Microvasc. Res.}, 2009, \textbf{77},
  265--272\relax
\mciteBstWouldAddEndPuncttrue
\mciteSetBstMidEndSepPunct{\mcitedefaultmidpunct}
{\mcitedefaultendpunct}{\mcitedefaultseppunct}\relax
\EndOfBibitem
\bibitem[Janoschek \emph{et~al.}(2011)Janoschek, Mancini, Harting, and
  Toschi]{janoschek2011rotational}
F.~Janoschek, F.~Mancini, J.~Harting and F.~Toschi, \emph{Phil. Trans. Roy.
  Soc. A}, 2011 \textbf{369}, 2337--2344\relax
\mciteBstWouldAddEndPuncttrue
\mciteSetBstMidEndSepPunct{\mcitedefaultmidpunct}
{\mcitedefaultendpunct}{\mcitedefaultseppunct}\relax
\EndOfBibitem
\bibitem[Pfafferott \emph{et~al.}(1985)Pfafferott, Nash, and
  Meiselman]{pfafferott1985red}
C.~Pfafferott, G.~Nash and H.~Meiselman, \emph{Biophys. J.}, 1985, \textbf{47},
  695--704\relax
\mciteBstWouldAddEndPuncttrue
\mciteSetBstMidEndSepPunct{\mcitedefaultmidpunct}
{\mcitedefaultendpunct}{\mcitedefaultseppunct}\relax
\EndOfBibitem
\bibitem[Saupe(1968)]{saupe1968recent}
A.~Saupe, \emph{Angew. Chem. Int. Ed.}, 1968, \textbf{7}, 97\relax
\mciteBstWouldAddEndPuncttrue
\mciteSetBstMidEndSepPunct{\mcitedefaultmidpunct}
{\mcitedefaultendpunct}{\mcitedefaultseppunct}\relax
\EndOfBibitem
\bibitem[Vlahovska \emph{et~al.}(2011)Vlahovska, Young, Danker, and
  Misbah]{vlahovska_dynamics_2011}
P.~M. Vlahovska, Y.-N. Young, G.~Danker and C.~Misbah, \emph{J. Fluid Mech.},
  2011, \textbf{678}, 221--247\relax
\mciteBstWouldAddEndPuncttrue
\mciteSetBstMidEndSepPunct{\mcitedefaultmidpunct}
{\mcitedefaultendpunct}{\mcitedefaultseppunct}\relax
\EndOfBibitem
\bibitem[Bagchi and Kalluri(2009)]{bagchi_dynamics_2009}
P.~Bagchi and R.~Kalluri, \emph{Phys. Rev. E}, 2009, \textbf{80}, 016307\relax
\mciteBstWouldAddEndPuncttrue
\mciteSetBstMidEndSepPunct{\mcitedefaultmidpunct}
{\mcitedefaultendpunct}{\mcitedefaultseppunct}\relax
\EndOfBibitem
\bibitem[Clausen and Aidun(2010)]{clausen_capsule_2010}
J.~R. Clausen and C.~K. Aidun, \emph{Phys. Fluids}, 2010, \textbf{22},
  123302\relax
\mciteBstWouldAddEndPuncttrue
\mciteSetBstMidEndSepPunct{\mcitedefaultmidpunct}
{\mcitedefaultendpunct}{\mcitedefaultseppunct}\relax
\EndOfBibitem
\bibitem[Barthès-Biesel(1980)]{barthes-biesel_motion_1980}
D.~Barthès-Biesel, \emph{J. Fluid Mech.}, 1980, \textbf{100}, 831–853\relax
\mciteBstWouldAddEndPuncttrue
\mciteSetBstMidEndSepPunct{\mcitedefaultmidpunct}
{\mcitedefaultendpunct}{\mcitedefaultseppunct}\relax
\EndOfBibitem
\bibitem[Hur \emph{et~al.}(2001)Hur, Shaqfeh, Babcock, Smith, and
  Chu]{hur_dynamics_2001}
J.~S. Hur, E.~S.~G. Shaqfeh, H.~P. Babcock, D.~E. Smith and S.~Chu, \emph{J.
  Rheol.}, 2001, \textbf{45}, 421\relax
\mciteBstWouldAddEndPuncttrue
\mciteSetBstMidEndSepPunct{\mcitedefaultmidpunct}
{\mcitedefaultendpunct}{\mcitedefaultseppunct}\relax
\EndOfBibitem
\bibitem[Huang \emph{et~al.}(2010)Huang, Winkler, Sutmann, and
  Gomppper]{huang_semidilute_2010}
C.-C. Huang, R.~G. Winkler, G.~Sutmann and G.~Gomppper, \emph{Macromol.}, 2010,
  \textbf{43}, 10107\relax
\mciteBstWouldAddEndPuncttrue
\mciteSetBstMidEndSepPunct{\mcitedefaultmidpunct}
{\mcitedefaultendpunct}{\mcitedefaultseppunct}\relax
\EndOfBibitem
\bibitem[Singh \emph{et~al.}(2012)Singh, Fedosov, Chatterji, Winkler, and
  Gompper]{singh_conformational_2012}
S.~P. Singh, D.~A. Fedosov, A.~Chatterji, R.~G. Winkler and G.~Gompper,
  \emph{J. Phys.: Cond. Mat.}, 2012, \textbf{24}, 464103\relax
\mciteBstWouldAddEndPuncttrue
\mciteSetBstMidEndSepPunct{\mcitedefaultmidpunct}
{\mcitedefaultendpunct}{\mcitedefaultseppunct}\relax
\EndOfBibitem
\bibitem[Chien(1970)]{chien_shear_1970}
S.~Chien, \emph{Science}, 1970, \textbf{168}, 977--979\relax
\mciteBstWouldAddEndPuncttrue
\mciteSetBstMidEndSepPunct{\mcitedefaultmidpunct}
{\mcitedefaultendpunct}{\mcitedefaultseppunct}\relax
\EndOfBibitem
\bibitem[Janoschek \emph{et~al.}(2010)Janoschek, Toschi, and
  Harting]{janoschek2010simplified}
F.~Janoschek, F.~Toschi and J.~Harting, \emph{Phys. Rev. E}, 2010, \textbf{82},
  056710\relax
\mciteBstWouldAddEndPuncttrue
\mciteSetBstMidEndSepPunct{\mcitedefaultmidpunct}
{\mcitedefaultendpunct}{\mcitedefaultseppunct}\relax
\EndOfBibitem
\bibitem[Melchionna(2011)]{melchionna_model_2011}
S.~Melchionna, \emph{Macromol. Theor. Simul.}, 2011, \textbf{20},
  548–561\relax
\mciteBstWouldAddEndPuncttrue
\mciteSetBstMidEndSepPunct{\mcitedefaultmidpunct}
{\mcitedefaultendpunct}{\mcitedefaultseppunct}\relax
\EndOfBibitem
\bibitem[Varnik \emph{et~al.}(2003)Varnik, Bocquet, Barrat, and
  Berthier]{varnik2003shear}
F.~Varnik, L.~Bocquet, J.~Barrat and L.~Berthier, \emph{Phys. Rev. Lett.},
  2003, \textbf{90}, 95702\relax
\mciteBstWouldAddEndPuncttrue
\mciteSetBstMidEndSepPunct{\mcitedefaultmidpunct}
{\mcitedefaultendpunct}{\mcitedefaultseppunct}\relax
\EndOfBibitem
\bibitem[Mandal \emph{et~al.}(2012)Mandal, Gross, Raabe, and
  Varnik]{mandal2012heterogeneous}
S.~Mandal, M.~Gross, D.~Raabe and F.~Varnik, \emph{Phys. Rev. Lett.}, 2012,
  \textbf{108}, 098301\relax
\mciteBstWouldAddEndPuncttrue
\mciteSetBstMidEndSepPunct{\mcitedefaultmidpunct}
{\mcitedefaultendpunct}{\mcitedefaultseppunct}\relax
\EndOfBibitem
\end{mcitethebibliography}
\end{document}